\documentclass[preprint]{aastex}

\begin{document}

\title{Preserving Structure in Multi-wavelength Images of Extended Objects}

\author{James Pizagno\altaffilmark{1}}

\altaffiltext{1}{jpizagno@u.washington.edu, Department of Astronomy, University of Washington, Seattle, USA}



\begin{abstract}
A non-parametric smoothing method is presented that reduces noise
in multi-wavelength imaging data sets.  Using Principle 
Component Analysis (hereafter PCA) to associate pixels according to their $ugriz$-band colors, 
smoothing is done over pixels with a similar location in PCA space.
This method smoothes over pixels with similar color, which reduces the
amount of mixing of different colors within the smoothing region.  
The method is tested using a mock galaxy with signal-to-noise levels and color
characteristics of SDSS data.  When comparing this method to smoothing 
methods using a fixed radial profile or an adaptive radial profile, 
the $\chi^2$-like statistic for the method presented here is smaller.  
The method shows a small dependence on input parameters.  
Running this method on SDSS data and fitting theoretical stellar population 
models to the smoothed data of the mock galaxy and SDSS data, shows that the method reduces scatter in the
best-fit stellar population analysis parameters, when compared to cases where
no smoothing is done.  For an area centered on the star forming region of the mock galaxy, 
the median and standard deviation of the PCA-smoothed data is 7 Myr ($\pm$ 3 Myr), as 
compared to 10 Myr ($\pm$ 1 Myr) for a simple radial average,  where the noise-free true 
value is 7.5 Myr ($\pm$ 3.7 Myr).
\end{abstract}


\clearpage

\section{Introduction}

Galaxy formation theories predict that baryons cool in dark matter halos in such 
a way as to provide connections between galaxy observables and dark matter properties \citep{whi78,col89}.
For example, the Tully-Fisher relation \citep{tul77} shows the connection between 
galaxy luminosity and circular velocity, where the circular velocity depends on the dark matter and baryonic 
mass profiles.  A key tool in studying galaxy evolution is the 
semi-analytic model approach to relating the observable
properties of galaxies to the underlying formation physics \citep{eis96,mo98}.
Semi-analytic models of galaxy formation predict observables
such as the luminosity function, radii, rotation curves, clustering 
statistics, colors, and the stellar mass of galaxies.  \cite{gne07} and \cite{dut07}  
have shown that semi-analytic models can predict the joint 
distribution of galaxy observables, with model parameters that depend on the 
baryonic mass profile, where the baryonic mass profile includes the stellar mass
and gas mass.  Budgeting baryonic mass into stellar mass and gas mass is an 
important tunable parameter in the modeling \citep{mcg05}.  One major difficulty lies 
in converting multi-wavelength imaging data into the stellar mass, where the multi-wavelength
imaging data is inherently noisy.  

Large surveys produce multi-wavelength maps of galaxies,
which are used to measure the baryonic properties of the galaxy 
population.  Large surveys produce large data sets,
which are comparable in size to modern simulations \citep{del06,bow06}.  
Since models of galaxy formation predict stellar mass and surface density, 
the multi-wavelength maps of galaxies must be converted into stellar 
populations using synthetic stellar population models \citep{mar05,bru03}.
In order to analyze the stellar populations in multi-wavelength
images of galaxies, noise-reduction techniques must be 
employed. Large surveys are key to answering these questions because
they provide data uniformity over a large area of the sky.  SDSS provides 
$ugriz$-band data over 25\% of the sky with a photometric calibration 
accuracy good to 2\% \citep{ive04}.  The existence of these noisy, but uniform and large, 
data sets, along with the need for stellar population modeling, requires a 
smoothing technique for multi-wavelength data sets.
 
A study similar to this one is (\cite{lan07}; hereafter L07), 
which studies the pixel color magnitude relation for nearby
galaxies.  L07 noted the distinct difference of pixel color magnitude 
diagrams with different Hubble types, where Early-type galaxies have redder 
pCDMs.  L07 noted how morphological features were related to distinct 
features in the pCDM.  Scatter in a pCDM was caused by extinction,
showing the need for accurate ISM extinction models when modeling
observed colors.  The study by L07 shows how pixel maps of 
galaxies are correlated with galaxy type, and might reveal hidden features.
L07 does not employ noise-reduction techniques, such as the ones presented
in this paper, which may affect the structure of the pixel diagrams.

Another study similar to this one is \cite{wel08}.
\cite{wel08} used the pixel-z technique, which combines stellar population
synthesis models with multi-wavelength pixel photometry of galaxies to study 
the stellar population content of SDSS galaxies.  \cite{wel08} showed how the star formation 
rate varies with local galaxy density, varies with position in a galaxy, and studied 
the mean star formation rate.  The pixel-z method does not include any smoothing
techniques.  This current work will be complementary to that work, by providing a 
 smoothing technique that will minimize the effect data noise will have on the best-fit
stellar population parameters. 

Adaptsmooth \citep{zib09} is a multi-wavelength smoothing algorithm that is similar to this work.  
Adaptsmooth uses a circularly symmetric radial median to reduce noise.
The radius of the circle is defined so that median smoothed data has a signal-to-noise of 20.  
All of the imaging wavelength data (i.e. $ugriz$-bands) are smoothed to the same  
radius, which is determined from the maximum radius of the multi-wavelength data in question, which 
is usually the $u$-band or $z$-band in SDSS data as they have the lowest signal-to-noise.  
Adaptsmooth is adaptive, in the sense that the radius varies with position in the galaxy, 
as the signal-to-noise varies. However, the radial median filter is still azimuthally symmetric.  
This means that blue star formation regions can get median filtered together with redder disk.
The PCA-smoothing method presented in this paper, median filters over pixels that are associated 
in PCA space according to their color.   

We focus on SDSS because it is a large and uniform data-set.  SDSS has coverage 
over 25\% of the sky, where the imaging data covers a large range of optical 
wavelengths, and does so in a uniform manner.  The distribution of galaxy properties
has been well studied for SDSS data sets \citep{bla03}.   Many semi-analytic models
have been constrained using SDSS data \citep{gne07,li07}.

In the following paper the technique is described in Section 2, a comparison to other methods is done 
in Section 3, case studies are presented in Section 4, and the conclusion is presented in Section 5.  

\section{Technique} 
The goal of this method is to smooth data without mixing different colors.  For example,
a simple radial smoothing method may mix a red bulge with a blue star forming region, 
which will result in poorly fit stellar population models.  The method must be 
automated so that it can be applied to large data sets, such as the SDSS, and not have 
input parameters that vary within the data-set.  Since the noise characteristics may vary 
among different data sets (i.e. SDSS vs. 2MASS), the method must also have tunable parameters.  
It is shown below that an advantage of this method is that it is not overly sensitive to the input 
parameters.  

The method presented here uses Principal Component Analysis (hereafter PCA) in order to 
quantitatively associate pixels according to their color.  PCA can be used 
to reduce the dimensionality of data, illuminate hidden correlations, quantify 
levels of proportionality, and rotate data into new axes.  
PCA has been used in many different areas of data analysis, and has been described in detail in 
other papers using it as an analysis tool for astronomical spectra \citep{yip04,con95}.  
PCA is used here as a way to associate pixels according to their color 
and spatial location.  This method transforms the multi-wavelength data, from 
color as a function of position in the galaxy to eigenweights of basis colors as a function of 
position in PCA space.  Poisson noise is reduced by averaging over pixels having 
similar PCA weights, or similar locations in PCA space.

\subsection{Mock Galaxy}
For several reasons, the method presented here is tested using a mock galaxy.  
First, the mock galaxy can be assigned a wide range of colors, representing 
those seen in real data sets.  The colors used here resemble the colors of recent star 
formation for the HII-like regions, old stellar populations for the Bulge, intermediate populations 
for the Disk, and dust is applied to different spatial regions.  Secondly, the effects of noise 
on different colors can be estimated by adding Poisson noise due to the source and background.  
Thirdly, the mock galaxy is assigned spatially variable colors with realistic spatial structures 
which may cause over simplified smoothing methods to fail.   For example, 
real galaxies tend to have a high signal-to-noise bulge with old reddish color, whereas 
the disk contains a fainter intermediate age color with a moderate-to-low signal-to-noise value, and
there are asymmetric HII regions with a young stellar populations that have high signal-to-noise values. 
Real galaxies have blue high SNR HII regions that are only a few FWHMs from a faint red dust lane.  Over 
simplified mapping techniques may average these colors together.
Fourthly, using a mock galaxy provides the ``truth''.  Knowing the true noise-free colors will allow a 
figure of merit to be calculated which can be used to test different techniques discussed in this paper.

The goal is to create a mock galaxy 
with realistic spatial variations of the underlying color and SNR values that are similar to SDSS data.
The components of the mock galaxy are a bulge, disk, spiral arms, star formation regions, and dust.
Using GALFIT \citep{pen02}, 5 different images are created, which represent the $ugriz$-band respectively.  
Each component has the color of a specific stellar population model.  Star formation is spatially 
modeled as an arc of point sources along the spiral arms.  A PSF, as a function of wavelength, was measured from real SDSS data, 
and was added to the GALFIT input file. The wavelength dependent extinction is applied 
to the $ugriz$-band mock images using the extinction curve in \citep{car89} and assuming a 
selective extinction value of $R_V=3.1$.  The PSF is then measured using stars at edge 
of the image, which were added by GALFIT as point-sources and the user-provided PSF.  
A mock galaxy was made that includes spatially varying colors, with dusty regions that are generated 
from low-pass filtering of a real galaxy. 

\begin{figure}
\plotone{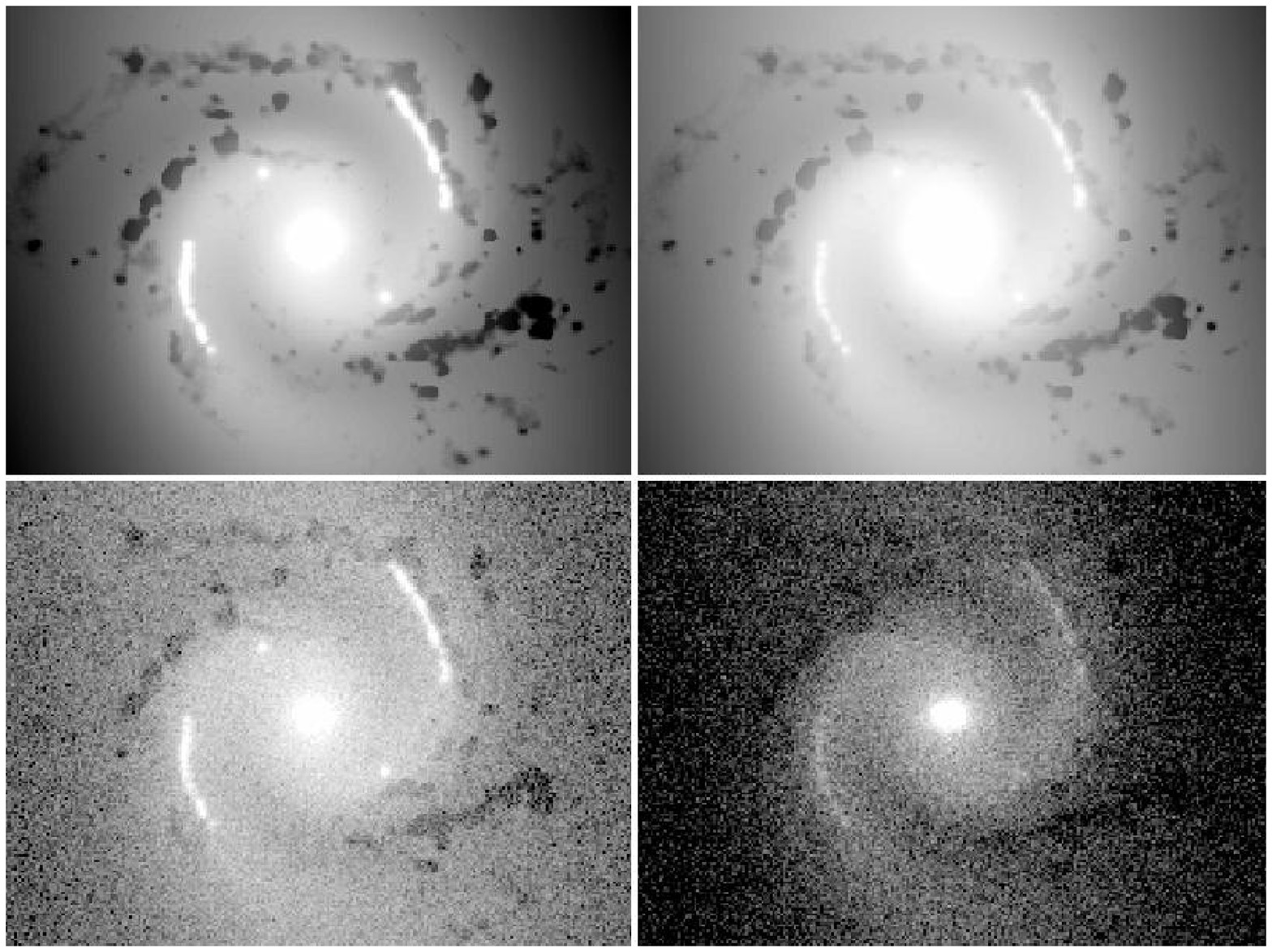}
\caption{The $g$-band and $r$-band images for the mock galaxy. The $g$-band image without noise 
(top left), $r$-band image without noise (top right), $g$-band image with noise (bottom left), $r$-band image with noise (bottom right).  
The $u,i,z$-band images are not shown.  The images were made using a combination of GALFIT models and 
real data.  
\label{fig:grmock}}
\end{figure}

Typical SDSS sky values for each band are added to the $ugriz$ images of the mock galaxy.
This is required to reproduce the SNR seen in SDSS data.  Poisson noise, including source
and sky, is added to the images of the mock galaxy.  The resulting $g$-band and $r$-band images 
are shown in Figure ~\ref{fig:grmock}.  Figure ~\ref{fig:grmock} shows the bulge, spiral arms, star formation regions, 
and dust regions.  These features qualitatively represent real SDSS data, in the sense
that the color varies with position and the features are asymmetric.  The mock galaxy matches the 
observed properties of SDSS galaxies with blue star formation regions and a red bulge, with a 
color of  $g-r=1.2$ for the bulge, $g-r=0.37$ for the exponential disk, $g-r=-0.63$ for the star formation regions,
and the patches of dust produce regions that are typically  $\Delta (g-r) \sim 0.12$ redder than the surrounding dust-free regions.  
Adding sky and source noise results in  total signal-to-noise values that vary from 50.0 in the center of the bulge
to 5-7 at the disk half-light radius, to nearly 0 where the outer parts of the disk fade into the sky background. 
For example, the bulge of the mock galaxy has an $r$-band flux $=$ 1500 DN on average, where the typical 
HII $r$-band flux $=$ 150-180, which is comparable to the galaxies from SDSS in Section 4.  

In the analysis steps below, the mock data is treated as if it were real SDSS data.  
PSFs are measured from point sources in the image.  The PSF of each image is matched
to the $u$-band PSF, because it is the broadest PSF.  The background sky was measured
from the corner of the images and subtracted from the data.  Before this method can be 
applied, the data must have PSFs matched, all instrument artifacts removed, and the galaxy identified. 

\subsection{PCA Application}
The method proposed here uses the results of PCA to determine a quantitative relationship
between the colors of different pixels, which are  associated during the smoothing
process.  It is assumed that each pixel is a linear combination of basis spectra.    
The flux at any pixel can be described as a linear combination of a set of weighted basis spectra.  
The normalized flux at pixel (x,y) can then be written as:
\begin{equation}
F_{\lambda} (x,y) = \sum_{i=1}^{5} a_{i}(x,y) e_{\lambda,i}
\end{equation}
where $i$ is the number of bands (5 for SDSS), $x$ and $y$ are spatial position in this galaxy, 
and $\lambda$ denotes the band (i.e. $ugriz$),  $a_i(x,y)$ is a eigenweight which varies as a function 
of position in the galaxy, and $e_{\lambda,i}$ is the $i$th basis at band $\lambda$.  

The covariance matrix method is used to measure the eigenvectors and eigenweights.  The data are first normalized to the $r$-band.
All pixels within 2 disk scale lengths and having a SNR ratio greater than a minimum value (discussed later) 
are included in a data matrix.  Using lower signal-to-noise values lower than this includes pixels heavily influenced by background sky colors.   
The covariance matrix of this data matrix is calculated, and then the eigenvectors 
and eigenvalues of this covariance matrix are determined.  This is carried out using Python procedures in the NUMPY.LINALG 
library, where the eigenvectors are solved using the APACK routines dgeev and zgeev \footnote{http://www.netlib.org/lapack/}.  

\begin{figure}
\plottwo{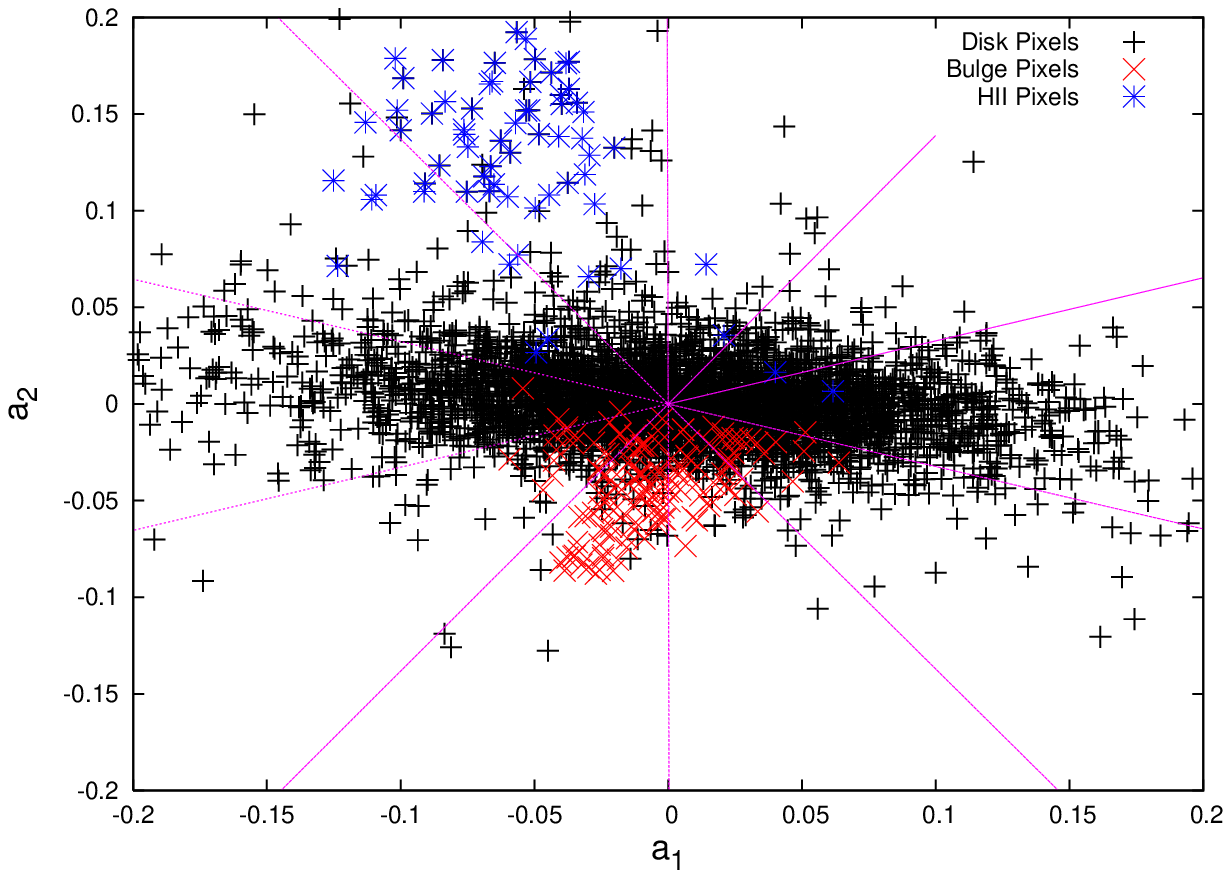}{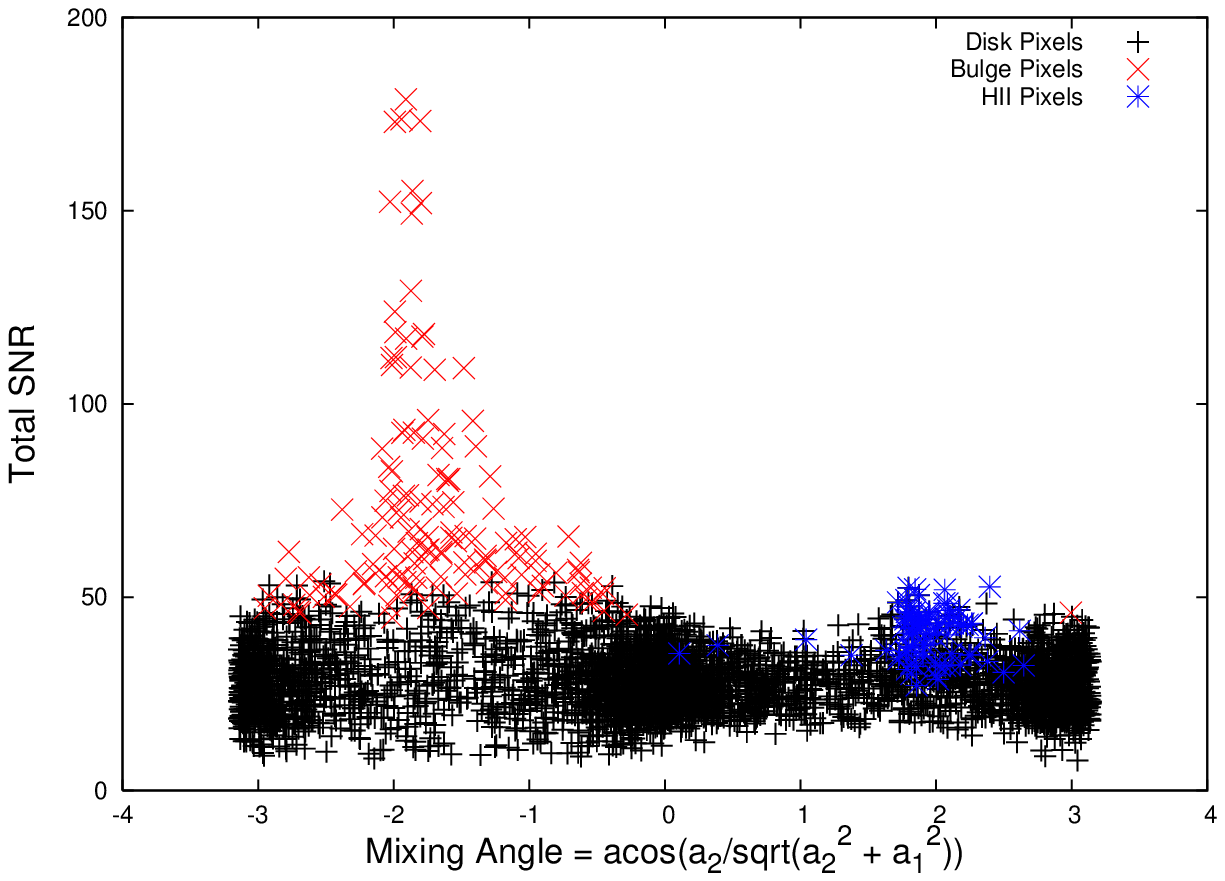}
\caption{The first and second eigencoefficients for the pixels for the mock galaxy. 
For display purposes, pixels are color-coded according to their spatial location in the galaxy.  
PCA is run on the galaxy using the pixels within 2 half-light radii and having a SNR greater than 3.  
The right panel shows, the total Signal-to-Noise ratio versus the angle in PCA-space.  
Points are color-coded according to spatial position in the galaxy.  
\label{fig:PCAanalysis}}
\end{figure}

Figure ~\ref{fig:PCAanalysis} shows the location of pixels within a subsection of the mock galaxy image.  The figure color-coding
is made according to spatial location in the galaxy. Pixels in similar 
areas of the left-hand panel of  Figure ~\ref{fig:PCAanalysis} have similar colors.  This method divides the PCA-space into evenly spaced 
angular bins.  Angular bins are chosen, as opposed to a linear interpolation, to follow the observed trend seen in 
PCA-space.  A small variation in color can result in a large variation in best-fit stellar population parameters.  
For example, a ($g$-$r$) color change of 0.3 can cause an estimate 
of the mass-to-light ratio to change by a factor of 3 (Figure 6 of \cite{bel03}).  Therefore,
the number of bins is chosen to have a standard deviation in $g$-$r$ color that will minimize the scatter in best-fit parameters.  
The number of bins used in smoothing is a free parameter, which can be adjusted by the user.  
The implementation here uses 10 bins because
using too few bins will produce averaging over different colors, whereas using too many 
bins will have bins that are so narrow that almost no averaging is done over pixels that are in 
similar areas of both PCA-space and spatial regions of the galaxy.  For example, using 5 bins 
breaks the PCA-space up so broadly that pixels in the same bin have a broad distribution 
with a standard deviation of $\delta (g-r) = 0.8$, whereas using 10 bins has a standard deviation of 
typically $\delta (g-r) = 0.3$.  To enhance the SNR, smoothing is done over pixels within a range of angular 
bins, where the range of bins is inversely proportional to that pixel's total signal-to-noise ratio.   More specifically,
for a pixel with total signal-to-noise $SNR(x,y)$, all pixels within the integer-value of $SNRenhanced/SNR(x,y)$ angular bins 
are median filtered together. For example, using an input parameter of $SNRenhanced=70$ for a pixel
with a $SNR(x,y)=30.0$, median filters within  $\pm$2 ($=int(70/30)$) bins in the left-hand panel of Figure ~\ref{fig:PCAanalysis}.
 The parameter $SNRenhanced$ is a constant for a given data set, and will be determined using the 
mock galaxy for the SDSS data set described here.  
This formulation has two effects. First, pixels with lower signal-to-noise values are averaged over more pixels.  
Secondly, the averaging is over pixels with similar colors in PCA space (Figure ~\ref{fig:PCAanalysis}).

A figure of merit metric ($FOM$) is used to quantitatively determine the best smoothing technique.  
It should be the difference between the truth and smoothed data, consider all wavelengths, and be inversely
proportional to the total noise value.  The $FOM$ is defined as:  
\begin{equation}
FOM(P,N) = \sum_{x,y}^N \sum_{\lambda=1}^5 \frac{(Truth_{\lambda}(x,y) - Smooth_{\lambda}(x,y|P))^2}{(\sigma_{\lambda}(x,y))^2},
\end{equation}
where $Truth_{\lambda}(x,y)$ is the true flux at band $\lambda$ at spatial position $(x,y)$, $Smooth_{\lambda}(x,y|P)$ is the
smoothed flux at band $\lambda$ at spatial position $(x,y)$,  $P$ is the list of free parameters ($SNRmax$, $SNRmin$, $Radius$, $SNRenhanced$), 
$\sigma_{\lambda}$ is the noise in band $\lambda$, and 
N is the number of pixels.  $SNRmax$ is the maximum SNR ratio of the pixels to which the method is applied.  $SNRmin$ is the 
minimum SNR over which smoothing is applied.  $Radius$ gives the size of the circle for pixels that may be included
in the mean.  $SNRenhanced$ is a constant controlling the range of pixels in PCA space that are included in the smoothing. 
The free parameters are then the ones mentioned above in $P$, along with the region within which the PCA eigen-vectors are 
measured,  and the number of PCA bins.  The best-fit parameters for the SDSS data-set 
are determined using a grid-search method, which searches for the parameters $P$ which provide the minimum 
$FOM$ .  The $FOM$ is defined in such a way that a lower $FOM$ is a better estimate of the true colors.

\begin{figure}
\plotone{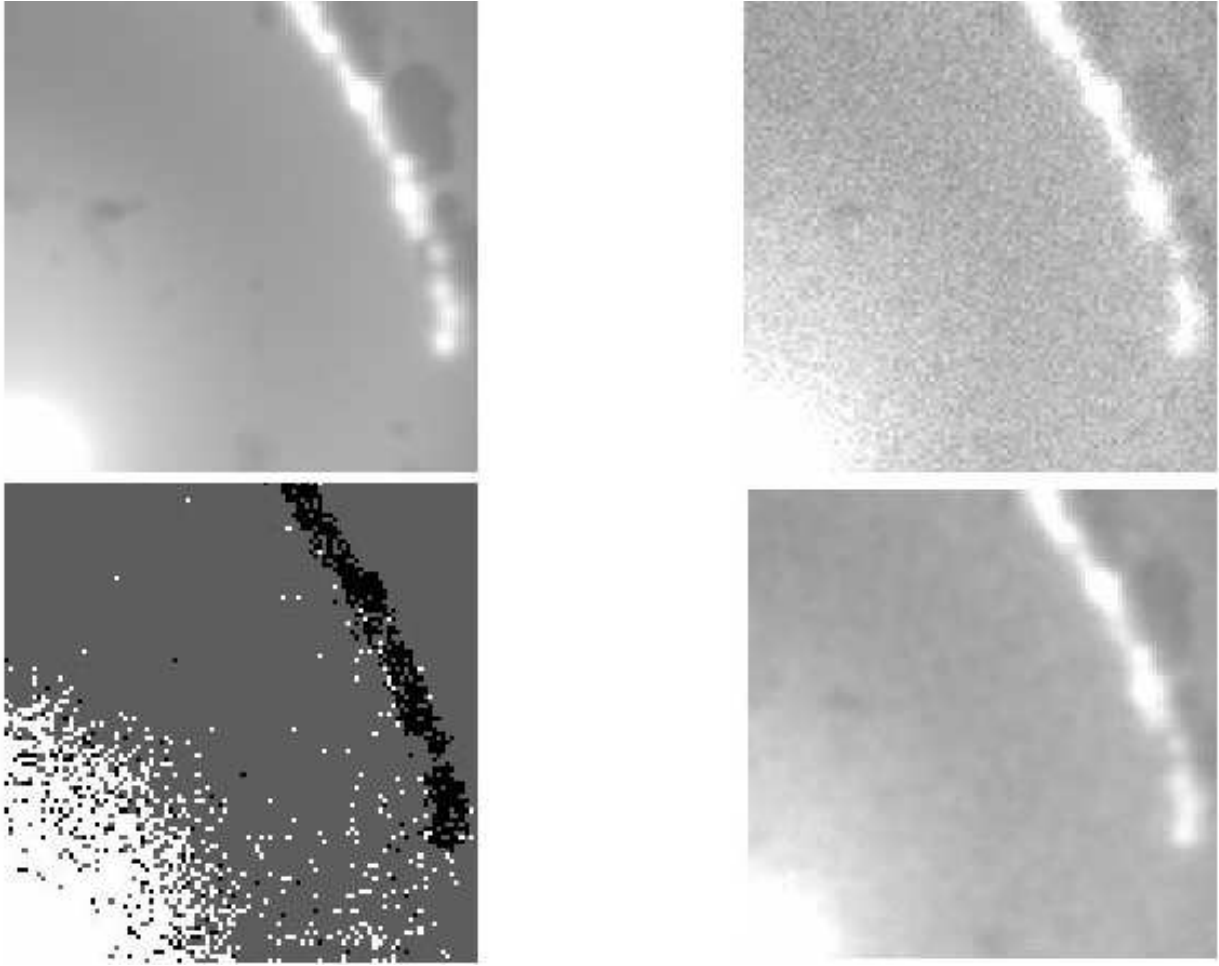}
\caption{The noise-free mock galaxy $g$-band image (top-left), noisy $g$-band image (top-right), PCA-map of pixels associated by SNR and angle in 
eigencoefficient space (bottom-left), and $g$-band image smoothed with the method presented here (bottom-right).   
The star-formation region and bulge are in different locations in PCA-space, as can be seen in the
PCA-map where the star-formation region is darker in color versus the lighter color bulge.  
\label{fig:PCAsmoothed}}
\end{figure}

Figure ~\ref{fig:PCAsmoothed} shows the region over which this experiment is run.   
The range of parameters searched was  $6<SNRmax<200$, $0<Radius<$ size-of-image, $6<SNRenhanced<600$, 
and $0<SNRmin<9$, and results are shown below for most of that range.  The bottom panels of Figure ~\ref{fig:PCAsmoothed}
 show the PCA map and smoothed PCA image.  The bottom left panel shows the combination of location of each pixel 
in the angular bin seen in Figure ~\ref{fig:PCAanalysis} and which pixels are included in smoothing.  The HII-like 
regions and bulge are clearly in different locations in PCA space, as can be seen by the fact
that they have different gray scales (angular bins).  The bottom right panel shows the image 
smoothed by pixels associated in PCA space (PCA-smoothing).  The contrast in the smoothed
image resembles the noise-free image (top left panel).  

\subsection{Analysis}
Figure ~\ref{fig:IMSTATall} shows the $FOM$ versus the free-parameters for the region in Figure ~\ref{fig:PCAsmoothed}.  
The lines show how the $FOM$ varies with the variable on the x-axis as the parameters in the key 
are held fixed.  The left panel of Figure ~\ref{fig:IMSTATall} shows how the $FOM$ varies as the maximum SNR value is varied.  
The $FOM$ has a minimum between $30 < SNRmax < 50$.  The reason the $FOM$ doesn't get smaller as SNRmax 
increases is because there are so few pixels with such high SNR values.  The middle panel of Figure ~\ref{fig:IMSTATall} 
shows how the $FOM$ varies with the smoothing radius.  For 
radii smaller than 3 the $FOM$ increases because less smoothing is done.  For radii larger than 4, 
the $FOM$ does not increase because only pixels with similar positions in PCA-space of
Figure ~\ref{fig:PCAanalysis} are included in the median.  The right panel of Figure ~\ref{fig:IMSTATall} 
shows the $FOM$ as the $SNRenhanced$ constant is changed.  This figure shows that any $SNRenhanced$ value greater 
than 40 will give an acceptable fit.  Increasing the $SNRenhanced$ value increases the number of bins over 
which pixels in PCA-space are smoothed.  The $FOM$ does not increase as $SNRenhanced$ is increased because only pixels
within a fixed spatial area (circle of radius 2-5) are included in the fit, and this mock galaxy does 
not have HII-like regions within a few pixels of the bright red bulge.  The $FOM$ does not change drastically as the parameters 
are changed slightly, demonstrating the robustness of PCA smoothing. 

Figure ~\ref{fig:ANALYZEHII} shows the $FOM$ versus the method parameters for a region centered on the HII-like
star forming arc.  The left panel of Figure ~\ref{fig:ANALYZEHII} shows how the $FOM$ varies with $SNRmax$.  The $FOM$ has a minimum 
between $20 < SMRNmax < 50$.  The middle panel of Figure ~\ref{fig:ANALYZEHII} shows how the $FOM$ varies
with $Radius$.  The $FOM$ has a minimum between $2< Radius< 4$.  The right panel of Figure ~\ref{fig:ANALYZEHII} 
shows how the $FOM$ varies with the $SNRenhanced$.  The SNR decreases rapidly to about 
$SNRenhanced=60$, and then does not decrease drastically.  It does not increase for this data because the HII-like region is not 
close to the red color bulge.  The variation of $SNRmin$ is not shown, because as long as it is $SNRmin=3$, there is an
acceptable fit.  These parameters, being 
only slightly different from the globally determined best-fit parameters, show the robustness of this method.  
Considering Figures ~\ref{fig:ANALYZEHII} and Figure ~\ref{fig:IMSTATall}, 
the best-fit parameters are SNRmax=30.0, Radius=4.0, SNRenhanced=60.0 and SNRmin=3.0.  

Figure ~\ref{fig:kernel} shows a map of pixels included in the  smoothing for a single pixel in the mock galaxy image.
The pixels with green crosses are included in the median average when smoothing for the central pixel.  The circle
is the best-fit radial aperture of $Radius=4$ pixels.  The pixels with green crosses all have similar locations in 
PCA-space (left panel of Figure ~\ref{fig:PCAanalysis}) and SNR range (right panel of Figure ~\ref{fig:PCAanalysis}).  
This figure shows that most of the pixels included in the median filter are HII-like region pixels.

Figure ~\ref{fig:SNRchange} shows the change in total SNR for pixels centered on the HII region.  
The SNR per pixel after PCA smoothing increases by a factor of 2-3 in the 
range of original signal-to-noise of 5-20.  Above the SNRlimit there is no change in 
the SNR. This is slightly less than the expected change in the SNR.  In a simple
circular average of radius=3 pixels, the SNR should increase by 5.3.
The SNR does not increase by a factor of 5.3 because not all pixels within the circular
aperture are used, as they are not in the same location in PCA space as the pixel be
being smoothed.

If the smoothing is not done over a range of angular bins in PCA-space (Figure ~\ref{fig:PCAanalysis}), or if 
$SNRenhanced=SNR(x,y)$ , then the $FOM$ always increases by 1.0.  The $FOM$ 
increases because there are so few pixels with the PCA-space in the same bin and within the same
spatial region.  
If the region is not restricted by a certain aperture, then the $FOM$ increases only slightly.
For example, if the radius is set to a value larger than the image, essentially including all
the pixels in the analysis, then the $FOM$ increases by only 0.8.  This is 
much better than the simple radial average, which increases by orders of magnitude when the radius 
is the size of the image.  
Changing the region over which the PCA eigenspectra are determined scatters the eigenvalues, reducing the
correlation between location in PCA-space and color.
Using a larger region includes pixels with such high noise values due to sky, so that PCA
results are scattered throughout the PCA-space.  Using too small a region does not include
a variation in color.  For example, using the central bulge half-light radius includes almost
no blue star formation colors.  

\section{Comparison with Other Methods}
We compare PCA smoothing with other smoothing techniques: a simple circular smoothing kernel 
and Adaptsmooth \citep{zib09}.  Adaptsmooth uses a circular aperture smoothing kernel, where 
the radius of the circle is set to achieve a SNR of 20.  
 
Adaptsmooth \citep{zib09} is compared to the PCA-smoothing technique for an 
area on the edge of an HII-like region of the mock galaxy.  Adaptsmooth is run 
in default mode, where the radial aperture is determined by increasing the radius until the 
resulting SNR equals 20.  Fixing the radius for all bands resembles the simple radial average results.  
Figures ~\ref{fig:IMSTATall} and ~\ref{fig:ANALYZEHII} show the $FOM$ for Adaptsmooth.  
In almost all cases the $FOM$ for Adaptsmooth is larger than the PCA-smoothing method presented here.
When the $FOM$ is calculated for the HII-like region, the results in ~\ref{fig:ANALYZEHII} shows that 
Adaptsmooth has a $FOM$ worse that no smoothing at all.  This is due to Adaptsmooth mixing pixels 
with different colors, which will be demonstrated below.  

Figure ~\ref{fig:Deltagr} shows a comparison between the ($g-r$) color predicted by various smoothing methods versus the 
true color, for a pixel located on the edge of the spiral arm.
For the simple circular smoothing kernel, an increase in the radius always produces a worse prediction 
of the color.  The color is off by 0.1 mag at all radii greater than 2 pixels.  Adaptsmooth 
picks a radius of 9 pixels for this pixel, as this was the maximum radius to reach a SNR=20 in the lowest
SNR band ($z$-band).  The radius of 9 pixels is so large that HII-like colors and redder disk colors
get mixed together in the median.  For PCA smoothing, with a best-fit $Radius=4$, the predicted $g-r$ color (open circle 
at $radius=4$pix is within 1-sigma of the true value (red filled square) and has a lower noise level.  Adaptsmooth (open triangle)
is more than 3 sigma away from the true value.   

The true $g$-band flux at the pixel in ~\ref{fig:Deltagr} is 152.782 DN.  Adaptsmooth predicts a 
$g$-band flux of 127.30 DN, whereas the PCA-smoothing predicts a more accurate flux of 157.93 DN.  The signal-to-noise
for this pixel is 10.85, where the PCA-smoothed SNR $=$ 28.63 versus the Adaptsmooth SNR$=$ 36.73.  There is clearly a trade off
when considering the number of bins to use ($SNRenhanced$), such that smoothing over more bins increases the SNR but decreases the
accuracy of the predicted color and vice-versa for using fewer bins.  The higher 
SNR for the Adaptsmooth result is due to over smoothing using too many pixels, which comes at the cost of a 
worse prediction of the SED ($g-r$ color in Figure ~\ref{fig:Deltagr}).  Figure ~\ref{fig:Deltagr} shows that the PCA-smoothing 
provides a better estimate of the true color, and has little dependence on the choice of radius. 

Next, the effectiveness of each method to reproduce the stellar population of the mock galaxy are discussed.  
\cite{mar05} stellar population models are fit to the true (noise-free)
mock galaxy image, mock galaxy image with no smoothing, a simple radially smoothed image, Adaptsmoothed image, and a 
PCA-smoothed image. The method uses a grid-search chi-squared minimization routine to find the best-fit model stellar 
population parameters, given the pixel's  $ugriz$-band fluxes and the model's $ugriz$-band fluxes.  For an aperture centered on the HII region, the 
noise-free truth image has an age of 7.5 Myr ($\pm$ 3.7 Myr).  The simple radial smoothing has an age of 10.0 Myr ($\pm$ 1.0 Myr).
Adaptsmooth results in ages of 6.0 Myr ($\pm$ 7.0 Myr).  The PCA-smoothed results in an age of 7.0 Myr ($\pm$ 4.1 Myr).  
The PCA-smoothed result is a better description of the noise-free true age.  

\section{Case Study}
\subsection{NGC 450}
Next PCA-smoothing is applied to real data sets. PCA-smoothing is first run on NGC 450, because
it was also analyzed by \cite{wel08}.  NGC 450 is a particularly interesting case 
because of the drastic variation in spatial colors.  There is a flocculent spatial distribution of 
very blue star formation regions on top of a red disk.  The PCA-smoothing was run on SDSS data of NGC 450 using the
best-fit parameters determined in Section 2. After running the PCA-smoothing, stellar population 
models \citep{mar05} are fit to the smoothed and un-smoothed data using a simple grid-search $\chi^2$ minimization 
routine.  Figure ~\ref{fig:ngc450color} shows the results.  Comparing the stellar population age maps with PCA-smoothing to unsmoothed maps shows 
that there is considerably more scatter in the un-smoothed maps.  Figure ~\ref{fig:ngc450color} also shows that the PCA-smoothing
preserves the structural information, where HII regions are still prominent in the PCA-smoothed image. 
The contrast between bright HII regions and a faint older disk is preserved during PCA-smoothing.

Figure ~\ref{fig:ngc450age2} quantifies the level of scatter in the best-fit ages.  Figure ~\ref{fig:ngc450age2} shows the 
analysis for a region with lower SNR pixels which are located on the disk region.  The scatter towards younger
ages for un-smoothed data is clear.  For unsmoothed data the best-fit ages range from 0.1 Gyr to 10 Gyr, where
the PCA-smoothed analysis has best-fit ages clustered around a few Gyr to 9 Gyr.  Since this is real data, the true age
distribution is not known and is not shown.  

\subsection{SDSS J235106.25+010324.1}
Next PCA-smoothing is run on SDSS J235106.25+010324.1, which has well defined blue spiral arms nearby a 
red bulge.  Figure ~\ref{fig:sdss} shows the SDSS color image, the PCA map, PCA-smoothed $g$-band image, best-fit model
age of the un-smoothed data, and best-fit model age of the smoothed data.  The top middle panel of Figure ~\ref{fig:sdss}
clearly shows that this method separates the bulge and spiral arms into different areas in PCA space, as can be seen by the 
bulge pixels being bright and the spiral arms being dark.  The separation of spiral arms versus 
the bulge and inter-disk regions means that they will not be mixed in the PCA-smoothing routines.  

\section{Conclusions}
This paper presents a method for smoothing SDSS data using a variation of Principal Component Analysis. 
The method is performed by running PCA simultaneously on multi-wavelength images of galaxies, and then 
smoothing over pixels that have similar locations in PCA space and spatial location within the galaxy.  
The advantages of the method are 1) no mixing of colors, 2) the method is geared towards stellar population analysis,  
3) the parameters are  tunable, and  4) the results are not extremely sensitive to the input parameters.  
The disadvantages of the method are 1) requiring initial analysis to identify the galaxy,  2) running PCA which may take computational time, 
and 3) requires well understood and uniform noise characteristic across different wavelengths.    
The smoothing parameters can be tuned to adjust the tradeoff between more smoothing and more color mixing versus less smoothing and 
more color purity.  Increasing the $SNRenhanced$ constant, results in an increased signal-to-noise of the smoothed pixel, at the cost of mixing over
different colors.  Lowering the $SNRenhanced$ constant, results in a more pure color with less smoothing over different colors, 
at the cost of a lower smoothed signal-to-noise.

\begin{figure}
\plotone{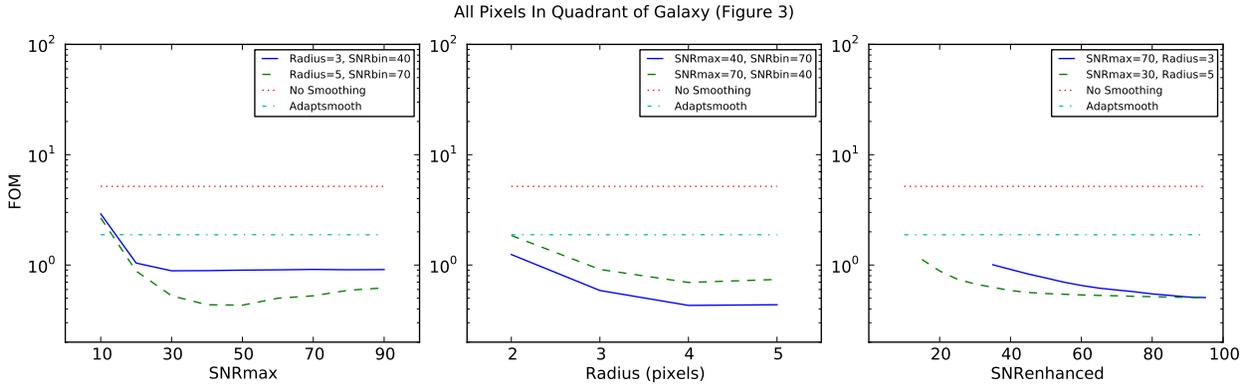}
\caption{
This figure show the $FOM$ versus the $SNRmax$, $Radius$, and $SNRenhanced$.  The $FOM$ for the PCA smoothed 
estimates (green-long dashed line, blue solid line) almost always has a lower $FOM$ when compared to 
Adaptsmooth.  This $FOM$ is calculated for all the pixels within the quadrant of the galaxy shown in Figure ~\ref{fig:PCAsmoothed}.  
The smoothly varying nature of the $FOM$ shows that this method is fairly robust to choice of input parameter.
These panels show acceptable $FOM$ values for Radii between 4-5 pixels, $30 < SNRmax < 50$, and $SNRenhanced > 40$.  
Adaptsmooth has a fixed value of $FOM$ in all the panels because Adaptsmooth is not a function of $SNRmax$ or $SNRenhanced$, and 
the radius varies with position in the galaxy.  For the region in Figure ~\ref{fig:PCAsmoothed}, Adaptsmooth uses different 
radii at different spatial location in the galaxy, which vary between 4 to 10 pixels.
\label{fig:IMSTATall}}
\end{figure}

\begin{figure}
\plotone{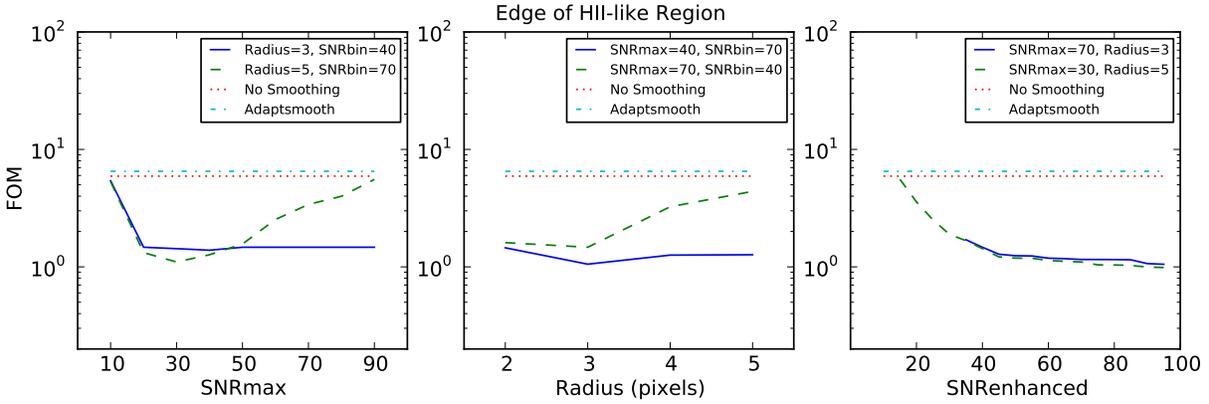}
\caption{The same as  Figure ~\ref{fig:IMSTATall}, except for a region centered on the edge of an HII-like region.
These panels show acceptable $FOM$ values for Radii between 2-4 pixels, $20 < SNRmax < 50$, and $SNRenhanced > 60$.  
Adaptsmooth has a fixed value of $FOM$ in all the panels because Adaptsmooth is not a function of $SNRmax$ or $SNRenhanced$, and 
the radius varies with position in the galaxy.  For this region, Adaptsmooth uses different radii at different position,
which vary between 9- 10 pixels.  
\label{fig:ANALYZEHII}}
\end{figure}

\begin{figure}
\epsscale{1.0}
\plotone{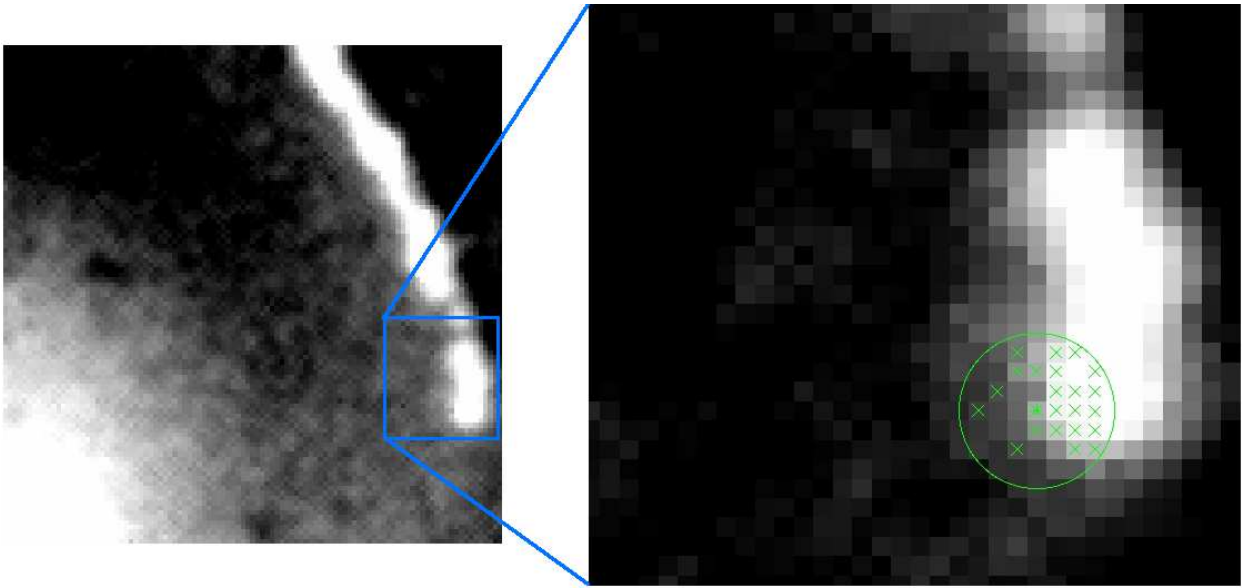}
\caption{A map of pixels that are included in an average (green cross) for selected pixels (center pixel, star).  
The pixel in the center will have its flux replaced by an average over the pixels labeled with a green cross.  
The circle shows the best-fit radial aperture of R=4 pixels.  Pixels with green crosses all have similar 
locations in PCA-space Figure ~\ref{fig:PCAanalysis}.  
\label{fig:kernel}}
\end{figure}

\begin{figure}
\epsscale{1.0}
\plotone{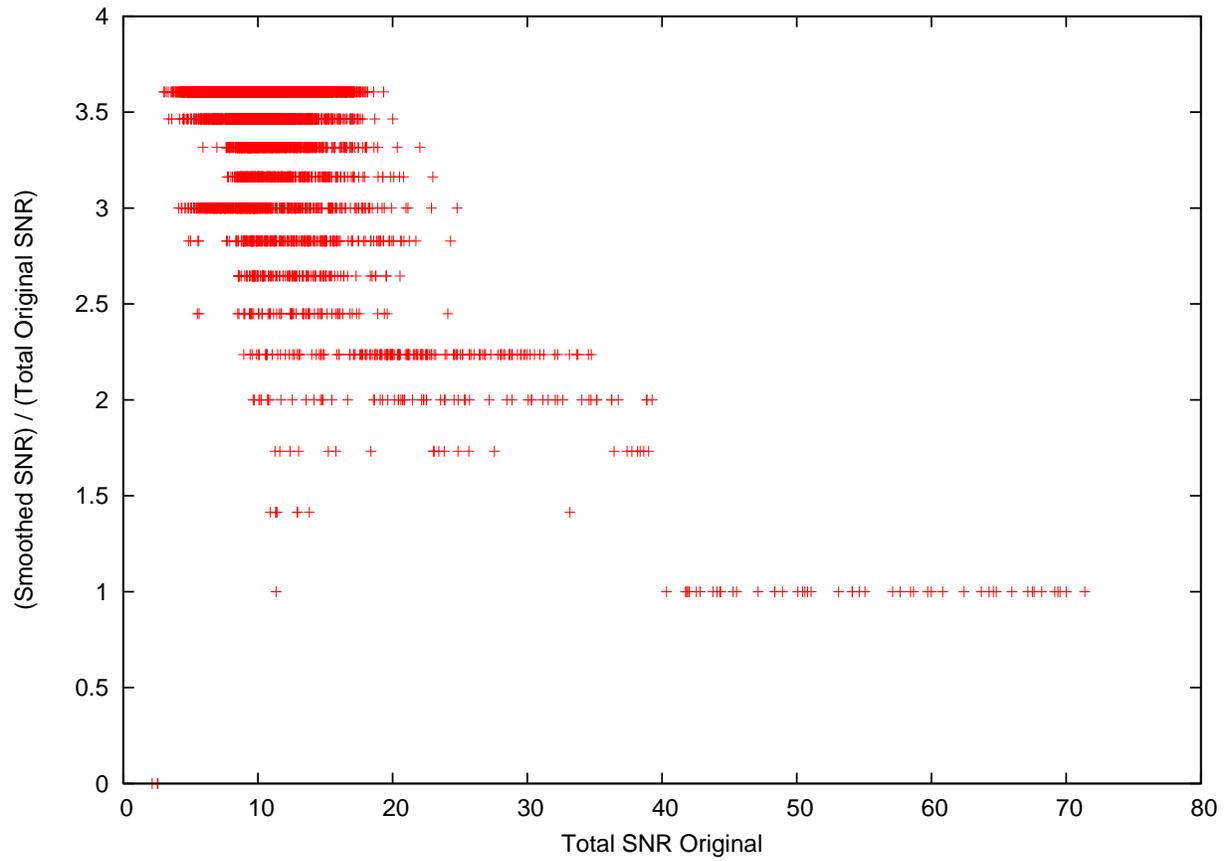}
\caption{The ratio of smoothed-to-original total SNR for the mock galaxy, for the best-fit parameters.  
This figure shows that this method provides more smoothing for pixels with lower SNR values.
\label{fig:SNRchange}}
\end{figure}

\begin{figure}
\plotone{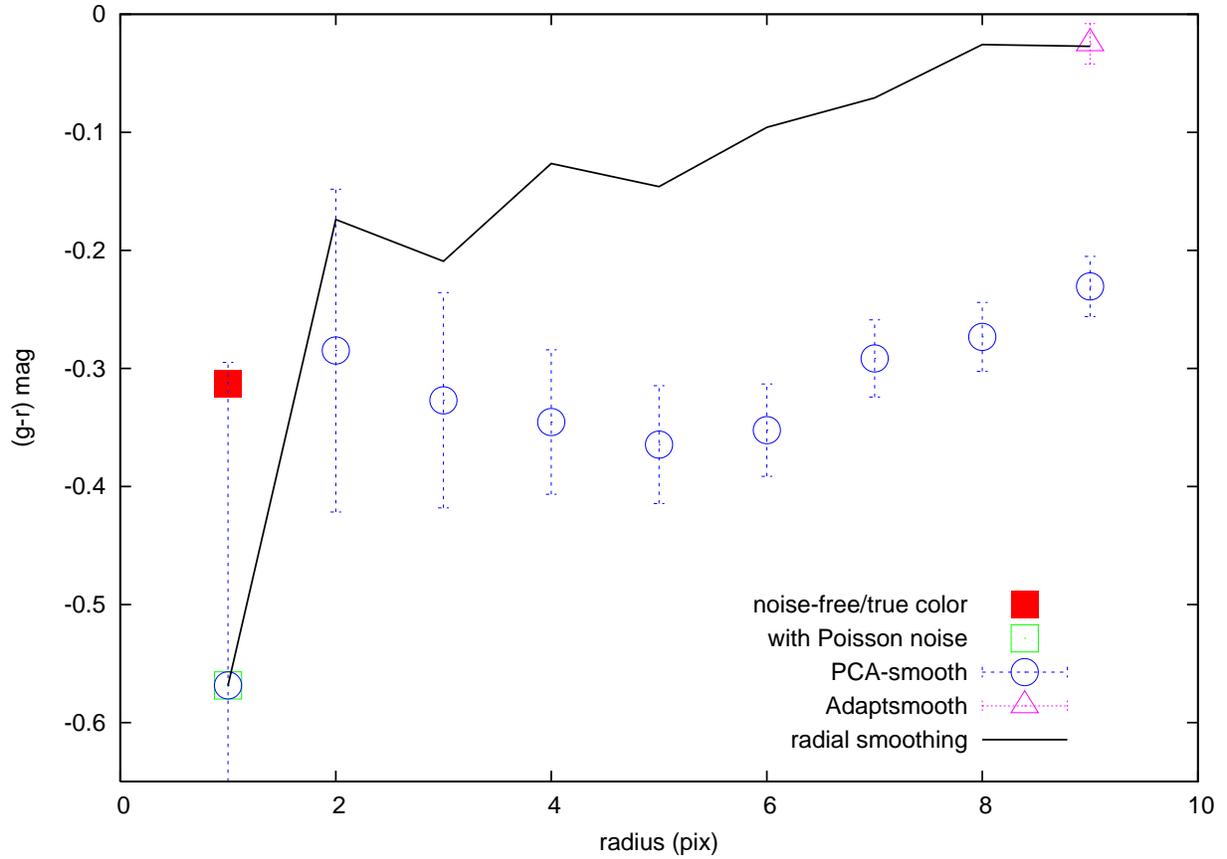}
\caption{This figure shows the comparison between different smoothing methods, for a single pixel located at the edge of the spiral arm.  
This figure shows the noise-free color of the mock galaxy or true color (solid red square), the color with Poisson noise due to sky and source added (green square), 
the PCA-smoothed color as a function of radius parameter (empty blue circles), the result of a Adaptsmooth (empty purple triangle), and the result of 
simple radial smoothing (solid black line).  
\label{fig:Deltagr}}
\end{figure}

\begin{figure}
\plotone{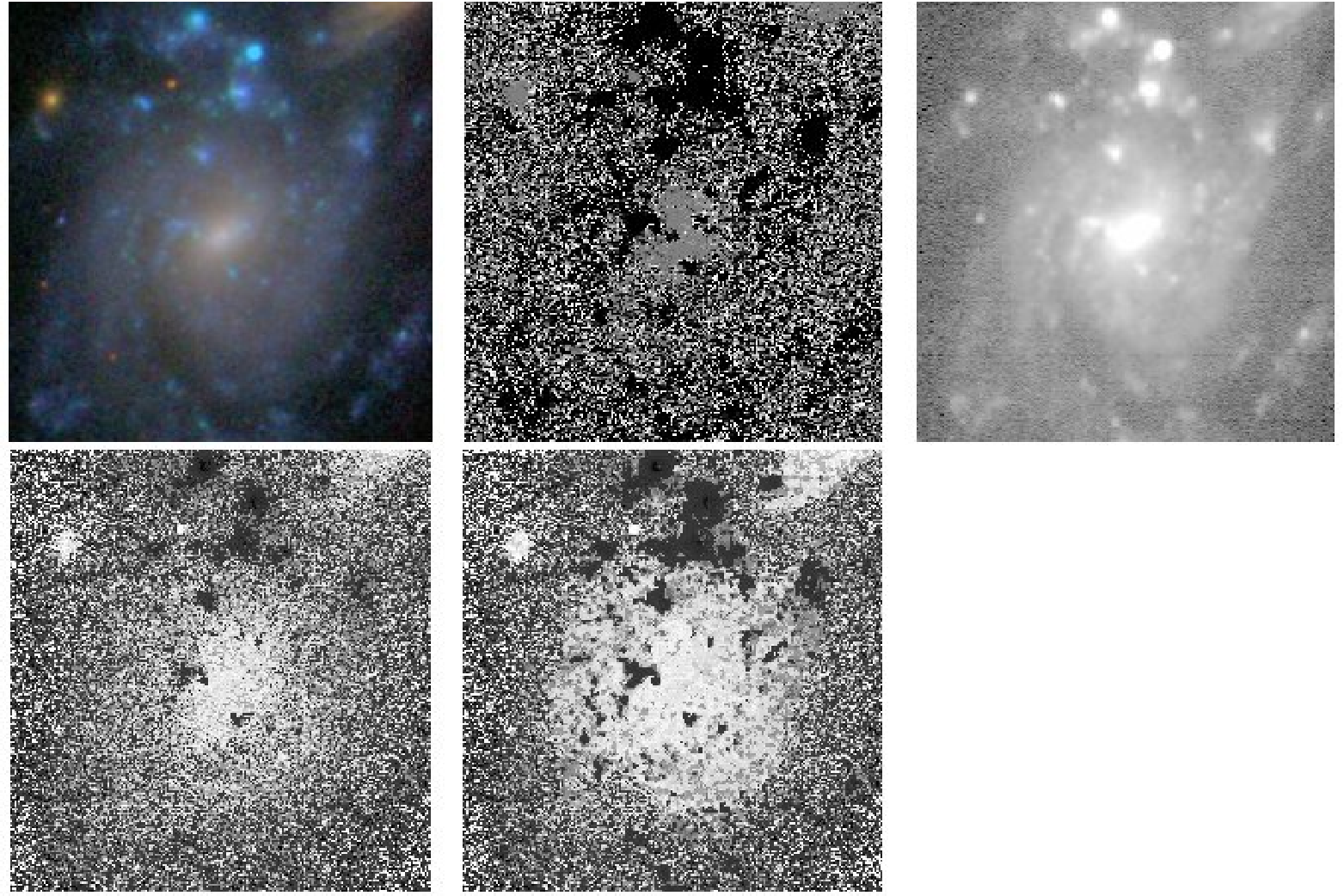}
\caption{A figure showing the SDSS color image (top left), the location in PCA-space (top center), the PCA-smoothed g-band image (top right), 
best-fit model age of the un-smoothed data (bottom left), and best-fit model age of the PCA-smoothed data (bottom right) 
for NGC450.  Comparison of the bottom two panels shows that there is qualitatively less scatter in the PCA-smoothed age. 
\label{fig:ngc450color}}
\end{figure}

\begin{figure}
\plotone{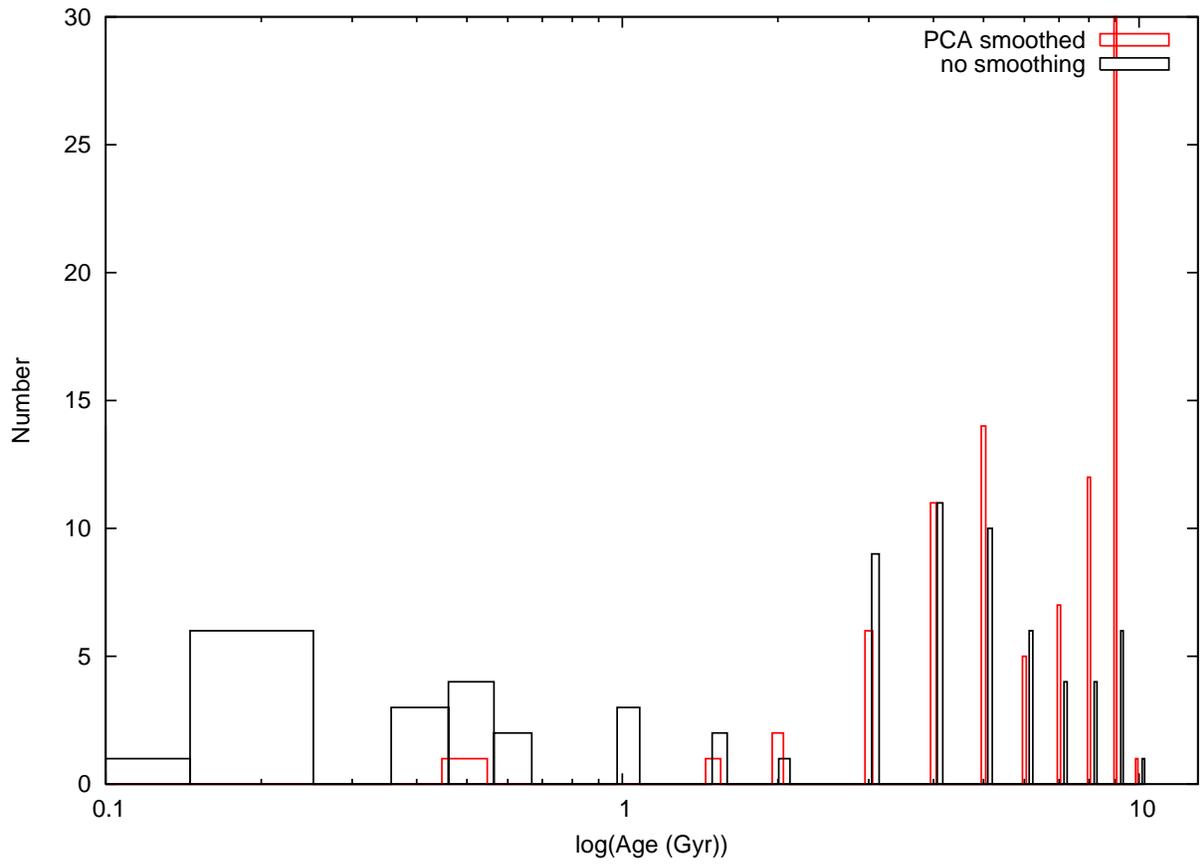}
\caption{A histogram of the best-fit Maraston model age for a low SNR region of NGC450. 
\label{fig:ngc450age2}}
\end{figure}

\begin{figure}
\plotone{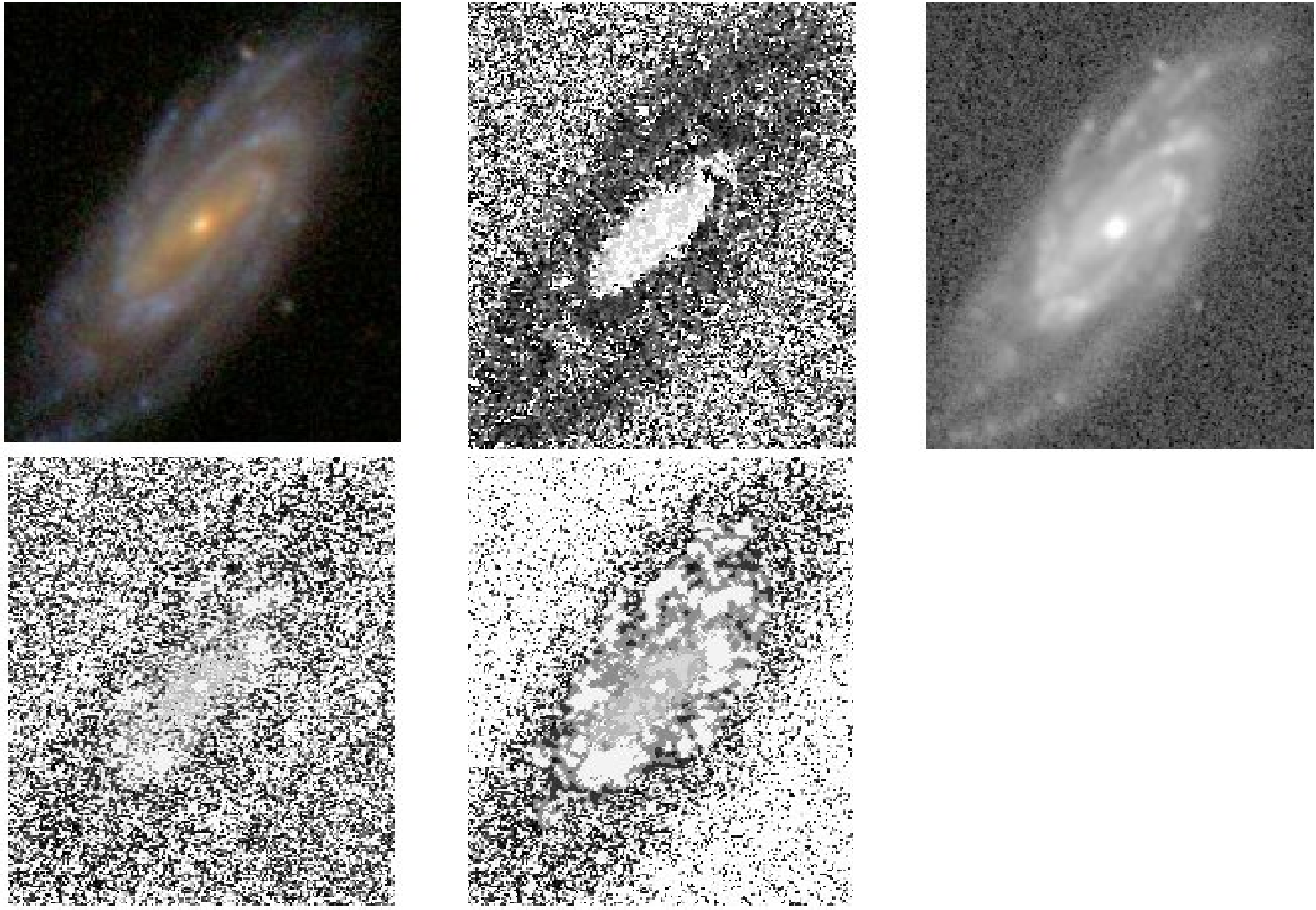}
\caption{A figure showing the SDSS color image (top left), the location in PCA-space (top center), the PCA-smoothed g-band image (top right), 
best-fit model age of the un-smoothed data (bottom left), and best-fit model age of the PCA-smoothed data (bottom right) 
for SDSS J235106.25+010324.1.  Comparison of the bottom two panels shows that there is qualitatively 
less scatter in the PCA-smoothed age. 
\label{fig:sdss} }
\end{figure}

The method was tested and demonstrated using a mock galaxy with $ugriz$-band images
having SNRs similar to that seen in typical SDSS data.  Figures ~\ref{fig:IMSTATall} and ~\ref{fig:ANALYZEHII} 
show that the $FOM$ for the PCA-smoothing method is always better (lower), 
when compared to azimuthally symmetric smoothing routines.  Considering Figures ~\ref{fig:ANALYZEHII} and Figure ~\ref{fig:IMSTATall}, 
the best-fit parameters are $SNRmax=30.0$, $Radius=4.0$, $SNRenhanced=60.0$ and $SNRmin=3.0$.   
The lack of extreme peaks in the $FOM$ shows the robustness of the method.  Figures ~\ref{fig:IMSTATall} and ~\ref{fig:ANALYZEHII} 
imply that as long as the user doesn't use extreme smoothing parameters, a reliable result will be obtained.  
Analysis of a region located on the boundary between an HII region and the red disk (Figure ~\ref{fig:Deltagr}), shows that PCA
smoothing is better at predicting a ($g$-$r$) color to 0.2 mag, when compared to simple radial smoothing or
Adaptsmooth.

The PCA-smoothing algorithm can be run on the SDSS data set, with the parameters described in this paper.  
The galaxies in the low-redshift NYU-VAGC \citep{bla05} would be perfect for analysis, as it includes galaxies 
within a comoving distance range of $10 < d < 150$ Mpc/h.  These nearby galaxies are spatially resolved, and 
perfect for this type of analysis.  This method is geared towards large area surveys having multi-wavelength data, 
over a large part of the sky, and having uniform noise characteristics (i.e. COSMOS, DEEP, SDSS, 2MASS, DES).  
The method can also be applied to Galactic Nebulae as well, which are also asymmetrical extended objects with 
multi-wavelength data.

\clearpage

\clearpage


\clearpage
\begin{thebibliography}{}

\bibitem[Bell et al.(2003)]{bel03}
Bell, E. F., McIntosh, D. H., Katz, N., Weinberg, M. D. 2003, ApJS, 149, 289

\bibitem[Blanton et al.(2003)]{bla03}
Blanton, M. R. et al. 2003, ApJ, 594, 186

\bibitem[Blanton et al.(2005)]{bla05}
Blanton et. al. 2005 MNRAS 129, 2562

\bibitem[Bower et al.(2006)]{bow06}
Bower, R. G., Benson, A. J., Malbon, R., Helley, J. C., Frenk, C. S., Baugh, C. M., Cole, S., Lacey, C. G. 2006, MNRAS, 370, 645

\bibitem[Bruzual \& Charlot (2003)]{bru03}
Bruzual, G., Charlot, S. 2003, MNRAS, 344, 1000

\bibitem[Cardelli, Clayton, Mathis(1989)]{car89}
Cardelli, J. A., Clayton, G. C., Mathis, J. S. 1989, ApJ, 345, 245

\bibitem[Cole \& Kaiser (1989)]{col89}
Cole, S., \& Kaiser, N. 1989, MNRAS, 237, 1127

\bibitem[Connolly et al.(1995)]{con95}
Connolly, A. J., Szalay, A. S., Bershady, M. A., Kinney, A. L., Calzetti, D. 1995, AJ, 110, 1071

\bibitem[De Lucia et al.(2006)]{del06}
De Lucia, G., Springel, V., White, S. D. M., Croton, D., Kauffmann, G. 2006, MNRAS, 366, 499

\bibitem[Dutton et al.(2007)]{dut07}
Dutton, A. A., van den Bosch, F. C., Dekel, A., Courteau, S. 2007, ApJ, 654, 27

\bibitem[Eisenstein \& Loeb(1996)]{eis96}
Eisenstein, D. J., \& Loeb, A. 1996, ApJ, 459, 432

\bibitem[Gnedin et al.(2007)]{gne07}
Gnedin, O. Y., Weinberg, D. H., Pizagno, J., Prada, F., Rix, H.-W. 2007, ApJ, 671, 1115

\bibitem[Ivezic et al.(2004)]{ive04}
Ivezi\'{c}, \v{Z}. et al. 2004, Astron. Nachr., 325, 583 

\bibitem[Lanyon-Foster et al.(2007)]{lan07}
Lanyon-Foster, M. M., Conselice, C. J., Merrifield, M. R. 2007, MNRAS, 380, 571

\bibitem[Li et al.(2007)]{li07}
Li, C., Jing, Y. P., Kauffmann, G., Boerner, G., Kang, X., Wange, L. 2007, MNRAS, 376, 984

\bibitem[Maraston (2005)]{mar05}
Maraston, C. 2005, MNRAS, 362, 799

\bibitem[McGaugh(2005)]{mcg05}
McGaugh, S. S. 2005, Phys. Rev. Lett., 95, 171302

\bibitem[Mo et al. (1998)]{mo98}
Mo, H. J., Mao, S., \& White, S. D. M. 1998, MNRAS, 295, 319

\bibitem[Peng et al.(2002)]{pen02}
Peng, C. Y., Ho, L. C., Impey, C. D., Rix, H.-W. 2002, AJ, 124, 266

\bibitem[Tully \& Fisher(1977)]{tul77}
Tully, R. B., \& Fisher, J. R. 1977, A\&A, 54, 661

\bibitem[Welikala et al.(2008)]{wel08}
Welikala, N., Connolly, A. J., Hopkins, A. M., Scranton, R., Conti, A. 2008, ApJ, 677, 970

\bibitem[White \& Rees(1978)]{whi78}
White, S. D. M., \& Rees, M. J. 1978, MNRAS, 183, 341

\bibitem[Yip, et al.(2004)]{yip04}
Yip, C. W., et al. 2004, AJ, 128, 2603

\bibitem[Zibetti, Charlot, Rix (2009)]{zib09}
Zibetti, S., Charlot, S., Rix, H.-W. 2009, MNRAS, 400, 1181

\end{thebibliography}
\end{document}